\documentstyle[aaspp]{article}

\newcommand{\bq}{\begin{equation}}
\newcommand{\eq}{\end{equation}}

\begin{document}
\def\refitem{\par\parskip 0pt\noindent\hangindent 20pt}

\title{A Preliminary Indication of Evolution of Type Ia Supernovae from their Risetimes}

\author{Adam G. Riess$^1$, Alexei V. Filippenko$^1$, Weidong Li$^1$,
and Brian P. Schmidt$^2$}

\affil{$^1$Department of Astronomy, University of California, Berkeley, CA 94720-3411,
$^2$Mount Stromlo and Siding Spring Observatories, Private Bag, Weston Creek, ACT 2611, Australia}

\begin{abstract}

   We have compared the risetime for samples
 of nearby and high-redshift type Ia supernovae (SNe Ia).
  The fiducial risetime of the 
nearby SNe Ia is 2.5$\pm0.4$ days longer than the preliminary measurement of the risetime by Goldhaber (1998a,b) and Groom (1998) for high-redshift SNe
 Ia from the Supernova Cosmology Project.
If further analysis of the high-redshift data does not lead to a significant change in the value or uncertainty in the risetime, then 
the statistical likelihood that the two samples have
 different fiducial risetimes remains high (5.8$\sigma$) and indicates possible
 evolution between the samples of SNe Ia.  We consider the likely
 effects of several sources of systematic error, but none of these
 resolves the difference in the risetimes.  
  Currently, we cannot directly determine the impact of the apparent
 evolution on previous determinations of cosmological parameters.  

\end{abstract}
subject headings:  supernovae: general$-$cosmology: observations

\vfill
\eject
 
\section{Introduction}

   High redshift (0.3 $< z <$ 1.0) type Ia supernovae (SNe Ia)
 are unexpectedly dim,
 a phenomenon readily attributed to a cosmological
   constant and an accelerating Universe (Riess et al. 1998; Perlmutter et
al. 1999).  These cosmological conclusions rely on the {\it assumption} that SNe Ia
 have not evolved.  Both the High-$z$ Supernova Search Team (Schmidt et al. 1998) and 
the Supernova Cosmology Project
(SCP; Perlmutter et al. 1997) have found no indication from spectra, light curves, and
various subsamples that SNe Ia have evolved between $z=0$ and $z=0.5$  
(Riess et al. 1998; Perlmutter et al. 1999); this evidence will be
considered in \S 4.  However, an unexpected luminosity evolution of $\sim$
   25\% over a lookback time of approximately 5
Gyr would be sufficient to nullify the cosmological conclusions.   Evolution
   is a notorious foe, plaguing previous
measurements of the global deceleration parameter using brightest
cluster galaxies (e.g., Sandage \& Hardy 1973).  
While we cannot hope to {\it prove} that the samples of SNe Ia
have not evolved, we would increase our confidence in their
reliability by adding to the list of ways in which they are similar
 while failing to discern any way in which they are different. 

  An important probe into supernova physics and possibly
  evolution is the rapid rise in luminosity of SNe Ia shortly after explosion.
  A wellspring of energy from birth, an expanding supernova releases ever
  dwindling resources of trapped energy from the radioactive decay of
  $^{56}$Ni and $^{56}$Co. The risetime (i.e., the time interval between explosion and
  peak) is dictated by the amount and location of synthesized $^{56}$Ni and the opacity of the
  intervening layers (Leibundgut \& Pinto 1992; Nugent et
al. 1995; Vacca \& Leibundgut 1996).

   Theoretical modeling indicates that expected variations in the
  composition of SN Ia progenitors at high redshift could be accompanied by
 an evolution in luminosity not accounted for by current empirical
  distance techniques (H\"{o}flich, Wheeler, \& Thielemann 1998).  One
  predicted signature of this evolution is an alteration of the risetime.  

   A preliminary measurement of the risetime using a large set of pre-discovery images of high-redshift ($z \approx 0.5 $) SNe Ia from the SCP was presented by Goldhaber (1998a,b) and Groom (1998).    Although the early SN Ia
 light is not strongly detected for individual objects, the statistics of $\sim$ 
40 different SNe Ia
have been used to meaningfully measure the fiducial risetime.  The final results from this measurement will be presented by Goldhaber et al. (1999).

  We have previously determined the risetime of nearby SNe Ia using a
  set of discovery and pre-discovery images from a mix of amateur and professional
  supernova searches (Riess et al. 1999).  A comparison of the high-redshift and low-redshift rise
behavior (\S 2) should provide a valuable test of evolution. 
  We discuss systematic errors which could bias
  this comparison in \S 3,  and the implications in \S 4.

\section{Analysis}

   Details concerning the acquisition and photometric calibration 
of very early observations of nearby SNe Ia can be found in Riess et
al. (1999).  These observations include $\sim$25 new measurements of
SNe Ia between 10 and 18 days before $B$ maximum.  The final analysis of the
high-redshift dataset and the details of the SCP's methods of analysis will be
presented in Goldhaber et al. (1999).  

   It is clear from the observations of nearby SNe Ia that there is 
   considerable inhomogeneity in the risetimes of SNe Ia (Riess et
   al. 1999).  However, individual SN Ia risetimes are well
   correlated with the post-rise light-curve shape (Goldhaber 1998a,b).  Therefore, one
   can determine both a
   fiducial risetime and the correlation between the risetime and 
the post-rise light-curve shape.  Riess et al. (1999)
   explored multiple techniques to quantify the fiducial SN Ia 
   risetime, all with consistent results. 

    To reliably compare the fiducial risetimes, it is essential to
   apply the {\it same method} of analysis to the high-redshift and low-redshift SN
   Ia datasets. 
For this reason, we emulated the stated methodology of Goldhaber (1998a,b) and Groom (1998) in analyzing the low-redshift data.
   
   We used the ``stretch'' method (Perlmutter et al. 1997) to
   normalize the $B$-band light curves of 10 nearby SNe Ia from Riess
   et al. (1999) using the same fiducial template, a modified Leibundgut (1989) template,  employed by
   Goldhaber (1998a,b) and Groom (1998).   (This template is very similar to the
   Leibundgut (1989) template.)   We performed this
   normalization using only observations later than
 10 days before maximum.  To emulate the lack of significant
   measurements of high-redshift SNe Ia at late times, we discarded
   observations past 35 days after $B$ maximum.
  The normalization parameters were then
   applied to the rise data (i.e., the observations earlier than 10 days
   before $B$ maximum).  As noted by Riess et al. (1999), this process
   results in an impressive decrease in the dispersion of the rise
   data from different SNe Ia.  

   After normalization, we fit the same empirical model proposed by
   Goldhaber (1998a,b) and Groom (1998) to the rise data.  This model is motivated
   by the description of a young SN Ia as a homologously expanding
   fireball whose initial luminosity is most sensitive to its changing radius (and relatively insensitive to the fractionally smaller changes in photospheric velocity and temperature).
   The luminosity is \bq L=\alpha (t+t_r)^2, \eq where
   $t$ is the time elapsed relative to maximum, $t_r$ is the risetime,
   and $\alpha$ is the ``speed'' of the rise. 

   The two free parameters, $t_r$ and $\alpha$, were determined by
   finding the best match between the model and data, i.e., when the 
standard $\chi^2$ statistic is minimized.  The minimum $\chi^2_\nu$
   was 0.82, indicating a good concordance between model and data.
   This fit and confidence intervals of the parameters are shown in Figure 1. 
Like Goldhaber (1998a,b) and Groom (1998), we identified the time of maximum as the
time when the SCP template reaches its brightest magnitude.
   (Uncertainties in determining the fiducial time of maximum do not
   affect a consistent comparison.)
 The resulting parameters from
   this fit were $t_r$=19.98$\pm0.15$ days and
   $\alpha$=0.071$\pm0.005$.  Because the SCP template is a good match
   to the Leibundgut (1989) template, it is not surprising that this risetime
   is quite similar to the value found by Riess et al. (1999) using
   the Leibundgut template as the fiducial template.  Besides the 29
   detections
 of SNe Ia between explosion and 10 days before maximum (relative to
   the SCP template), Riess et
   al. (1999) also provide 4 non-detection limits of SNe Ia 
in the temporal vicinity
   of explosion.   Unfortunately, these non-detections provide
   negligible additional constraints; in all cases they are more than
   3 mag above the expected luminosity of the SNe Ia (based on
   the fit to the detections).  

As seen in Figure 1, the post-rise
low-redshift data are well fit by the same modified Leibundgut (1989) template used to normalize the high-redshift data, verifying that the two data
sets have indeed been normalized to the same light-curve shape.  The light curve shown for the high-redshift SNe Ia prior to 10 days before maximum is the best fit of equation (1) to the preliminary SCP data (Goldhaber 1998a,b).  This model fit
clearly departs from the low-redshift data as seen in the residuals from the fit to the nearby rise in Figure 2.  By
13.8 days before maximum, the difference is 0.5
mag.  At 15.5 days before maximum, the difference rises to 1 mag.  

   The low-redshift risetime of 19.98$\pm0.15$ days is significantly longer than the
preliminary measurement of the risetime of 17.50$\pm0.40$ days found for high-redshift SNe Ia from the SCP
(Goldhaber 1998a,b; Goldhaber et al. 1999 will present final results).  The statistical likelihood that these risetime
measurements are discrepant is greater than 99.99\% (5.8$\sigma$).  

\section{Biases and Systematic Errors}

\subsection{Low Redshift}

  Most of the earliest observations of nearby SNe Ia were recorded with unfiltered CCDs and transformed to the $B$
  band.  
Riess et al. (1999) describe in detail a number of
  systematic tests they performed to appraise the influence of the
  transformation process on the estimate of the fiducial risetime.
  Employing different methods to calibrate the observations onto the standard passband system had little effect on the inferred
  risetime.  In addition, the risetime was found to be 
insensitive to the shape or
  color of the young SN Ia spectral energy distribution.  A comparison between transformed magnitudes and coeval
  magnitude measurements observed through standard passbands shows
  excellent agreement with no evidence for systematic departures.

  However, the discrepancy between the low-redshift and high-redshift risetimes
is independent of {\it any} systematic uncertainties in the transformation of unfiltered CCD observations to a
standard passband.  The reason is that the two earliest (in the
``dilated'' timeframe of the SCP template) unfiltered SN
Ia detections, SN 1998bu and SN 1997bq,  
were recorded nearly one full day {\it before} the
explosion time inferred from the SCP data (see Figure 3).  
Moreover, it is not possible
to detect SNe Ia outside the Local Group of galaxies less than
$\sim$0.5 days after explosion with small-aperture telescopes (in this
case the Beijing Astronomical Observatory and the amateur
 telescope of C. Faranda).  Therefore, we conclude that
the fiducial risetime of the low-redshift SNe Ia must be {\it at least} 1.5 days
longer than the preliminary risetime measurement inferred by Goldhaber (1998a,b) and Groom (1998) for the high-redshift
SNe Ia, independent of the reliability of photometric
transformations described in Riess et al. (1999).  This difference alone is significant at the
99.99\% (3.8$\sigma$) confidence level.

   Riess et al. (1999) discuss the possibility that an intrinsic
   dispersion in risetime (for a given light-curve shape) 
   can lead to the preferential inclusion of slowly rising SNe in
   our nearby sample and a bias in the inferred risetime.  Although
   the best-fit $\chi^2_\nu$ does not support additional 
intrinsic dispersion, Riess et al. (1999) considered a
   subsample of SNe Ia whose membership is independent of the rise.
  From the unbiased set we infer a risetime of 20.42$\pm0.34$ days,  inconsistent with the SCP
 preliminary measurement of the risetime at the 99.99\% (5.6$\sigma$) confidence level.

  The difference in risetimes does not seem to be a result of the stretch method.
  Even if this method were to distort the true risetime, this
 should not affect the risetime {\it comparison} because
 the light-curve shapes represented in the nearby sample span
 the range of light-curve shapes of the SCP sample.  The SCP sample of SN Ia light curves has a narrow distribution of stretch factors around unity
 (Perlmutter et al. 1999).  The
 average light-curve shape for our sample is 94\%$\pm$9\% of the mean width
 of the SCP light curves.  SN Ia light curves in the nearby sample have stretch
 factors which are both smaller and larger than unity.

\subsection{High Redshift}

    Correctly measuring the rise of faint 
SNe Ia at high redshift is a
    great technical challenge which must be convincingly overcome before we can
trust the implications of the comparison to the low-redshift rise
behavior.  The differences in the rising curve of the low-redshift and
high-redshift SNe Ia are only significant at 12 days before $B$
maximum and younger, when SNe Ia are more than 2 magnitudes below their peak
brightness.  For SNe Ia with redshifts of 0.4 to 0.5, this corresponds
to observed $R$-band magnitudes of 24.0 to 24.3 (Garnavich et al. 1998), which are 
$K$-corrected to $B$ magnitudes of 24.8 to 25.1.
Even larger differences between the low and high-redshift rise are evident when the SNe Ia are 3 mag below maximum, requiring
observations of high-redshift SNe Ia at $R$-band magnitudes of 25.0
to 25.3.  These faint fluxes push the limits of what can be
accomplished with 4-m-class telescopes under reasonable conditions and
moderate integrations.  Many of these individual 
observations of SNe Ia 
at high redshift have a signal-to-noise ratio near unity, a regime where
careful data analysis is required to avoid systematic errors in photometry.  

   The SCP observations of high-redshift SNe Ia on the rise (i.e., those 10 to 25 days before $B$ maximum) 
originate from reference images.
  These are observations taken 3 to 4 weeks before
a subsequent set of ``search'' observations and are used to measure host
galaxy brightnesses without SN light.
SNe Ia found during the
search phase are preferentially discovered near maximum brightness.  Due to
time dilation, the reference observations therefore contain the light
of SNe $\sim$14 to 18 days before maximum.  The SCP has taken great care to
obtain ``final'' reference images years after (or before) the SN Ia
explosions to accurately assess the amount of SN Ia light in the original
reference images (Perlmutter et al. 1999).
  They also employ light curves of SNe Ia to determine the amount of any residual light remaining in
the final reference images (Perlmutter 1999).  Consequently, we expect the SCP measurement of the risetime of
high-redshift SNe Ia to be accurate and comparisons to the
low-redshift risetime to be meaningful.  

   However, a very powerful crosscheck of systematic errors on the SCP's faint
   photometry of young high-redshift SNe Ia comes from examining their
   data of SNe Ia well past maximum when the SNe Ia are again of similar
   brightness.  In the age range of 35 to 45 days past maximum,
   the nearby SNe Ia are 2.7 to 3.1
   magnitudes below their maximum brightness.  This is the same flux
   range at which differences of 0.6 to 0.8 mag are evident in
   the similarly normalized low and high-redshift rising SN Ia light curves.
   Figure 4 shows a comparison of the low and high-redshift behavior
   of SNe Ia (normalized to the composite light curve of the high-redshift SNe Ia) on the rising and
   declining sides of maximum at identical magnitudes.  {\it This
   comparison demonstrates a high degree of concurrence of the declining
   light curves at the same magnitude levels where the disparity
   occurs on the rising light curves.}  The difference in the mean between the
   high-redshift and low-redshift magnitudes on the decline and in this
   magnitude range is less than 0.02 mag, indicating that
   conspicuous systematic errors in the faint SCP photometry cannot explain the
   differences in the rise behavior.  It is important to note that the
   data in the age range of 35 to 45 days past maximum were not used in
   the process of normalizing the light curves, making this an
 independent test.

   As discussed in \S 3.1, an intrinsic dispersion in SN Ia risetime
   (for SNe Ia with the same post-rise light-curve shape),
   together with a selection criterion related to the brightness of
   SNe Ia on the rise, could
   also bias the high-redshift measurement.  Because
   high-redshift SNe Ia are discovered by their appearance in
   differenced images, SNe Ia which are fainter than average in the
   reference observations will display a larger change in the
   ``search'' images.  This effect would seem to favor the discovery
   of SNe Ia which are faster or have shorter risetimes than
   the average for a given light-curve shape.  However,
 the individual measurement uncertainties in the SCP
   reference observations are larger than the differences
   in the low-redshift and high-redshift rise (Goldhaber 1998a,b).
  Therefore the
   criterion used to discover a SN Ia, the signal-to-noise ratio of the change
   in flux, is unaltered by changes in flux at this level.  
Because SNe Ia in the high-redshift sample and the unbiased subset
of the nearby sample were not discovered during the early rise, the apparent
difference in the samples' risetimes is not simply a result of a
selection bias.  

The transformation of observed high-redshift SN Ia magnitudes
   to rest-frame magnitudes requires the application of
   cross-filter $K$-corrections (Kim, Goobar, \& Perlmutter 1996).  Systematic errors
   in these corrections are likely at the 0.03 to 0.05 mag level (Perlmutter et
   al. 1999; Kim et al. 1996; Schmidt et al. 1998) but cannot explain
   the observed differences of $\sim$0.7 mag at 14-15 days before $B$
   maximum.

Additional sources of systematic errors in the SCP
   measurement of the SN Ia high-redshift risetime are best considered
   by Goldhaber et al. (1999).
   
    Despite the above tests, it would be desirable to have an
   additional crosscheck in the form of an independent measurement
   of the high-redshift risetime. Although the data of
 the SCP are currently the most extensive available on the rise of high-redshift SNe Ia,
early detections were made by the High-$z$ team of SN 1995K at $z=0.48$
 (Schmidt et al. 1998) and SN 1996K at $z=0.38$ (Riess et al. 1998). 
Unfortunately, with only two data points it is not
   possible to derive an independent, meaningful comparison to the
   risetime measurements.  More early rise data are needed
    from the High-$z$ team to yield an accurate measurement of the risetime.

\section{Discussion}

    Our measurement of the risetime from nearby SNe Ia is inconsistent
   with the preliminary measurement of the risetime of high-redshift SNe Ia 
inferred by the SCP (Goldhaber
   1998a,b, Groom 1998; see Goldhaber et
   al. 1999 for final results) with high
   statistical confidence (5.8$\sigma$).   The sense of the difference is that 
the
   low-redshift risetime measurement is 2.5$\pm0.4$ days longer than the
   high-redshift measurement.  This difference 
must be either the result
   of a systematic error in the measurements or intrinsic to SNe Ia.
  No compelling source of
   systematic error was found in \S 3 which could
   bring the low-redshift and high-redshift risetimes into concordance.
   However, systematic errors in the preliminary high-redshift measurement are
   best addressed by Goldhaber et al. (1999).  

We attempted to follow the methodology of Goldhaber (1998a,b) and Groom (1998) in 
making this comparison, and
{\it if} they have correctly followed their stated methodology,
we believe the difference reported here is significant.  This difference can 
only be rendered insignificant if future analysis of the high-redshift data 
concludes that a substantial error was made in determining the high-redshift 
risetime {\it or its uncertainty}.  The significantly higher quality of the 
low-redshift photometry and the existence of strong early detections of nearby 
SNe Ia makes it highly unlikely that a systematic error in the low-redshift 
risetime measurement could alone bring the disparate risetimes into concordance. 
A further comparison
between the data sets cannot be made at this time due to the unavailability of 
the SCP photometry.

It is surprising, considering the relatively low signal-to-noise ratio of the 
high-redshift SN Ia photometry, that a measurement of the risetime could be made 
to within the stated precision (1$\sigma$ = 0.4 days; Groom 1998).  
Indeed, further careful 
analysis of the high-redshift data could result in a larger measurement 
uncertainty.  For example, if the precision of the high-redshift risetime 
measurement decreased significantly to 1$\sigma$ = 1 to 1.5 days, the 
significance of the difference in the measured risetimes would reduce to 
only $\sim$95\% (i.e., 2$\sigma$). 
If further analysis indicates an extreme increase in the uncertainty of the 
high-redshift risetime measurement, 
the current precision of the comparison of the risetimes as 
a test of evolution might 
become insufficient to reach a robust conclusion.

       If the difference in the risetime behavior is intrinsic to the
    SNe Ia, this would likely indicate an evolution of the
    characteristics of SN Ia explosions.  Synonymous with evolution 
is the existence of a previously unknown, additional parameter not
    included in the current one-parameter empirical description of SN Ia 
light curves whose typical value evolves with redshift.
What would be the
    implications of an evolution of SNe Ia between $z=0$ and $z=0.5$? 

   An evolution of SNe Ia may reveal a redshift-dependent 
variation in the composition of SN Ia progenitors (Ruiz-Lapuente \& Canal 1998; Livio et al. 1999, Kobayashi et al. 1998). 
The construction of SN Ia progenitors is limited by the time required
for stars to reach their degenerate endstates and for transfer of
sufficient material from the donor star.  Because white dwarfs
born from low-mass stars will be absent at high redshifts, some
evolution of SNe Ia may be expected. The apparent evolution of SNe Ia
may augment our currently limited understanding of the nature of SN Ia
progenitors.

   SNe Ia have also been employed as a powerful tool for measuring cosmological
parameters.  In this role, the observed faintness of high-redshift SNe Ia has
been taken as evidence for a current acceleration of the Universe due
to a cosmological constant (Riess et al. 1998; Perlmutter et al. 1999).  If SNe Ia are evolving, how are previous measurements of
cosmological parameters from SNe Ia affected? 

The observed evolution of SNe Ia during their
 rise could only impact distance measurements if this evolution
 extends to the post-rise development.  Only observations of the brightness
 and colors of SNe Ia near peak and a
 few weeks thereafter have been used to estimate their distances.   

   The observation of different risetimes for SNe with similar
   subsequent light-curve shapes may signal a departure of the
   viability of previous one-parameter empirical models.
   Unfortunately, we cannot
    yet {\it directly} infer
    the size or direction of the effect on the cosmological parameters.

  A pure empiricist could be guided simply by Ockham's razor to conclude
 that the two unexpected characteristics of
    high-redshift SNe Ia (that they appear to rise more quickly and to be
    systematically dimmer than expected)
 are most economically explained by a single
 hypothesis: they have evolved.  Such evolution might be expected
    between high and low redshifts where variations in metallicity and
    progenitor ages must occur.  This hypothesis would obviate the
 need for a cosmological constant.

  However, as noted by Schmidt et
 al. (1998), Riess et
 al. (1998), and Perlmutter et al. (1999), the sample of nearby SNe Ia
 already spans an impressive range of environments and stellar
 populations.  SNe Ia hosted by early-type, late-type, and starburst
 galaxies show no systematic differences in their distance estimates.
 The relative reliability of SN Ia distance measurements across
 expected variations in progenitor properties in the nearby Universe 
is arguably the best evidence that evolution since $z\approx0.5$ should
 not affect cosmological measurements.  Yet it has not been determined
 if the {\it specific} environments of progenitors in different host
 galaxy types vary substantially.  The individual environments of SNe
 Ia in the nearby Universe must be investigated before we can infer
 whether or not their {\it assumed} variability provides evidence
 against evolution.   It is important to note that none of the host
 galaxies used to measure the low-redshift risetime are early-type
 galaxies.  The complete distribution of host galaxy types used to measure the
 high-redshift risetime is not yet known (Perlmutter et al. 1999).

  Semi-empirical methods used to calibrate the maximum
luminosity of SNe Ia suggest that the peak luminosity may be affected
by
a change in the risetime, but without additional information these methods give
opposite indications of the direction of the change in peak luminosity
(Nugent et
al. 1995).  Treating a SN Ia as an expanding photosphere with all
other
variables unaffected, a shorter risetime results in a dimmer SN Ia at
peak.  Alternately, a determination of the peak luminosity from the
instantaneous rate of radioactive energy deposition indicates that
a shorter risetime yields a brighter peak.  Because the expanding
photosphere determination of the peak luminosity is a steeper function
of the risetime, the methods together suggest that a shorter
risetime yields a somewhat dimmer peak (Nugent et al. 1995).

Because the perceived differences are only apparent when the
    SNe Ia are very young, we might conclude that physical
    differences exist only on the surface or outer layer of the
    SNe.  This conclusion resonates with the observation that
    spectroscopic differences among normal and overluminous SNe Ia are most apparent at
    early times (e.g., Filippenko et
al. 1992; Phillips et al. 1992; Li et al. 1999).
  These superficial differences may be related to an
    aspect of the material recently accreted (but not thoroughly
    processed) onto the progenitor. 
    If this parameter were the metallicity of the
    accreted material, it would not be surprising that the opacity and
    hence the risetime could be affected.  A lower surface metallicity, expected at high redshift,
 would produce a faster or shorter risetime, in concordance with our
results.  However, modeling indicates that the photosphere of
a SN Ia may recede below the surface layer of unprocessed material
in only a few days (H\"{o}flich 1999; see also Lentz).    

If the only source of the observed risetime difference is the surface of the SN Ia,
 how would the peak luminosity be affected?  As a fraction of the peak output, the difference in
the total energy lost during a short or long rise is negligible.
If the conditions necessary for explosion are dictated by the 
progenitor mass or properties near the center, the variations in the surface chemistry would 
not affect the size of the energy source (i.e., the $^{56}$Ni mass).
Once the photosphere
    receded beneath the surface, the subsequent evolution of the SN Ia
    including the peak luminosity may be unaltered.

  Because the risetime is a function of the diffusion time of
energy from the decay of $^{56}$Ni to the surface, the observed
difference in risetimes could indicate a variation in the initial
location of the synthesized $^{56}$Ni.  Longer risetimes would result
from SNe Ia which had a greater depth of intermediate-mass elements
covering the $^{56}$Ni.  If this variation is caused only by
mixing, the diffences in the diffusion times at peak should be
negligible resulting in little or no variation in peak luminosity (Pinto 1999).

  Yet, it is also possible that the
    risetime difference could be indicative of a deeper evolution of the SN Ia
    explosion which is only observable at the surface where the
    unburnt material resides.    If the change in the risetime results from a more complex
 alteration of the SN Ia physics, we cannot easily infer the effect on the post-rise light-curve. 

   We might hope to employ detailed modeling of SN Ia explosions to
   gauge the effect that the observed evolution has on measurements of
   the cosmological parameters.  However, the inability of current theory to
    adequately model many of the observed characteristics of SNe Ia
 engenders little faith that theory alone can be used
    to predict the consequences of the observed evolution.
    Specifically, the value of the fiducial risetime and the trend
    between risetime and peak luminosity (or decline rate) 
is in poor concordance with
    most available theoretical models (Riess et al. 1999).

Theoretical models have
indicated that differences in white dwarf carbon-to-oxygen (C/O) ratios should produce
variations in SN Ia explosions (H\"{o}flich, Wheeler, \& Thielemann 1998).  
The similarity between stellar and cosmological timescales leads to
the conclusion that the white dwarfs which produce high-redshift SNe
Ia originate, on average, in younger and hence more massive stars than today's SNe Ia (von Hippel, Bothun, \& Schommer 1997).
  Variations in the C/O ratio may be a natural consequence of 
white dwarfs which evolved from different stellar masses.  

  H\"{o}flich et al. (1998) predict that
    decreasing the progenitor's C/O ratio by 60\% produces a SN Ia
    with the same decline rate, yet is 30\% brighter and requires 3 days longer to reach maximum
    brightness.  This, if low-redshift SNe Ia have significantly
    smaller C/O ratios than high-redshift SNe Ia, the direction and size of this effect obviates
    the need for a cosmological constant or an accelerating Universe
    to explain the observations of low and high-redshift SNe Ia. However, H\"{o}flich et al. (1999) also expect that more massive stars would yield
 white dwarfs with lower C/O ratios.  At higher
    redshifts, lower mass stars have not yet had time to become white
    dwarfs.  Therefore, the theoretical prediction would be for higher
    redshift progenitors to give rise to more slowly rising SNe Ia, which
 is opposite to the observed trend reported here.  In addition,
 others suggest an inverse relation to that of H\"{o}flich et al. (1998) between the C/O ratio and luminosity (Umeda et al. 1999).  
More work is needed to
    understand this complicated process.  

      Observations at high and low redshifts can directly test for
    risetime evolution and its cosmological implications.  An exploration of the rise
    behavior of nearby SNe Ia born in a wide range of environments may
    reveal objects more similar to those at high redshift;  by
    determining the relative luminosity of ``fast'' and ``slow'' rising SNe Ia
    in
the nearby Universe we could directly evaluate the impact of the
    apparent evolution on cosmological inferences.  Comparisons
    of the spectra of high and
    low-redshift SNe Ia observed during the early rise, for example,  may indicate
    systematic differences.  A measurement of the rise behavior of
    SNe Ia at $z\approx0.2$ should yield results which are between
    those presented here and the preliminary SCP measurement.   Finally,
    the most challenging but potentially most fruitful way to explore
    the role of evolution in the current cosmological measurements is
    by extending the measurement of high-redshift SNe Ia to $z
    > 1.0$, where the effect of luminosity evolution is likely to
    diverge from that of a vacuum energy density.  As seen in Figure 5, a simple linear luminosity evolution of SNe Ia 
mimics the effects of a cosmological constant and mass density only in a specific redshift range.
Continued degeneracy between evolution and cosmology at additional redshifts
can only be envisioned by the most imaginitive (and sadistic) minds.
 
We are indebted to Mark Armstrong, Eric Thouvenot and Chuck Faranda for 
providing the CCD images of young SNe Ia.
We wish to thank Ed Moran, Peter Meikle, Peter Nugent, Gerson Goldhaber, Saul Perlmutter, Don
Groom, Robert Kirshner, Peter Garnavich, Saurabh Jha, Nick Suntzeff and
Doug Leonard for helpful discussions.  The Aspen Center for Physics
provided a stimulating environment in which these results were
discussed.  The work at U.C. Berkeley was supported by the Miller Institute for Basic Research
in Science, by NSF grant AST-9417213, and by grant GO-7505 from
the Space Telescope Science Institute, which is operated by the
Association of Universities for Research in Astronomy, Inc., under
NASA contract NAS5-26555.

\vfill \eject
 
\centerline {\bf References}

\refitem Filippenko, A.V. et al. 1992, ApJ, 384, L15

\refitem Garnavich, P., et al. 1998, ApJ, 493, 53

\refitem Goldhaber, G., 1998a, B.A.A.S., 193, 4713

\refitem Goldhaber, G., 1998b, in ``Gravity: From the Hubble Length to
the Planck Length ,'' SLAC Summer Institute (Stanford, CA: Stanford
Linear Accelerator Center)

\refitem Goldhaber, G., et al., 1999, in preparation

\refitem Groom, D. E., 1998, B.A.A.S., 193, 11102

\refitem H\"{o}flich, P., Wheeler, J. C., \& Thielemann, F. K., 1998,
ApJ, 495, 617

\refitem H\"{o}flich, P., et al., in preparation

\refitem Jha, S., et al., 1999, ApJSS, in press (astro-ph/9906220)

\refitem Kim, A., Goobar, A., \& Perlmutter, S. 1996, PASP, 108, 190

\refitem Kobayashi, C., Tsujimoto, T., Nomoto, K., Hachisu, I., Kato,
M., 1998, ApJ, 503, 155

\refitem Leibundgut, B. 1989, PhD thesis, University of Basel

\refitem Leibundgut, B., \& Pinto, P. A., 1992, ApJ, 401, 49

\refitem Lentz, E. J., Baron, E., Branch, D., Hauschildt, P. H, \& Nugent, P. E., 1999, ApJ, submitted (astro-ph/9906016) 

\refitem Li, W., et al., 1999, AJ, in press

\refitem Livio, M., 1999, astro-ph/9903264

\refitem Nugent, P., Branch, D., Baron, E., Fisher, A., \& Vaughan,
T., 1995, PRL, 75, 394 (Erratum: 75, 1874)

\refitem Perlmutter, S., 1999, private communication 

\refitem Perlmutter, S., et al., 1997, ApJ, 483, 565

\refitem Perlmutter, S., et al., 1999, ApJ, 517, 565

\refitem Phillips, M. M., Wells, L., Suntzeff, N., Hamuy, M.,
Leibundgut, B., Kirshner, R. P., \& Foltz, C. 1992, AJ, 103, 1632

\refitem Pinto, P., 1999, private communication 

\refitem Riess, A. G., et al., 1998, AJ, 116, 1009

\refitem Riess, A. G., et al., 1999, AJ, submitted

\refitem Ruiz-Lapuente, P., \& Canal, R. 1998, ApJ, 497, 57

\refitem Sandage, A., \& Hardy, E., 1973, ApJ, 183, 743

\refitem Schmidt, B. P, et al. 1998, ApJ, 507, 46

\refitem Umeda, H. et al., 1999, astro-ph/9906192

\refitem Vacca, W. D., \& Leibundgut, B., 1996, ApJ, 471, L37

\refitem von Hippel, T., Bothun, G. D., \& Schommer, R. A. 1997, AJ,
114, 1154

{\bf FIGURE CAPTIONS:}

{\bf Fig 1.}-$B$-band data of nearby SNe Ia normalized by the ``stretch" method
to the same modified Leibundgut (1989) template used to normalize the high-redshift SNe Ia (Goldhaber 1998a,b) and the inferred risetime
parameters.  The observation times of the individual SNe Ia are dilated to provide the
best fit of the post-rise (i.e., after 10 days before maximum) data
({\it diamonds})
 to the modified Leibundgut (1989) template ({\it asterisks}).  Fitting a quadratic
rise model to the nearby rise data ({\it filled circles}) 
yields confidence intervals for the fiducial speed and risetime to
$B$ maximum.  The preliminary best-fit of the same model to the high-redshift data is shown as a dashed line.
The best estimate of risetime of nearby SNe Ia is 19.98$\pm0.15$ days, 
significantly longer and statistically inconsistent with the
preliminary measurement of the risetime
of high-redshift SNe Ia of 17.5$\pm$0.4 days (Goldhaber 1998a,b). 

{\bf Fig 2.}-Residuals from the best fit rise model of the nearby SNe Ia.
Individual observations of nearby SNe Ia are shown as filled circles
and the preliminary best model fit to the high-redshift data is shown as a dashed line.

{\bf Fig 3.}-Pre-explosion, post-explosion, and maximum light observations of SN 1998bu.
After dilating to match the modified Leibundgut template, the detection by
Faranda of SN 1998bu (middle panel)
is 18.5 days before $B$ maximum and 1 day {\it before} the explosion time 
expected from the high-redshift SNe Ia.  The existence of this detection is
strong evidence that the risetime to $B$ maximum is {\it at least} 1.5 days greater
than the preliminary value inferred by Goldhaber (1998a,b) and Groom (1998) for high-redshift SNe Ia. The image at maximum light is from Jha et al. (1999)

{\bf Fig 4.}-Comparison of nearby and high-redshift SNe Ia
at similar magnitudes below maximum on the rise and decline.  The excellent agreement
between the two samples on the decline is strong evidence that systematic errors
incurred in measuring faint SNe at high redshift on the rise is not the cause of the
apparent difference between fits to the rise of the samples.

{\bf Fig 5.}-The degeneracy between simple linear evolution and cosmological parameters
on the Hubble diagram of nearby and high-redshift SNe Ia.  The possible confusion between
the effect of evolution and a cosmological constant on SN Ia distances could be resolved
by additional measurements of SNe Ia at redshifts greater than one.
The data shown are from Riess et al. (1998).

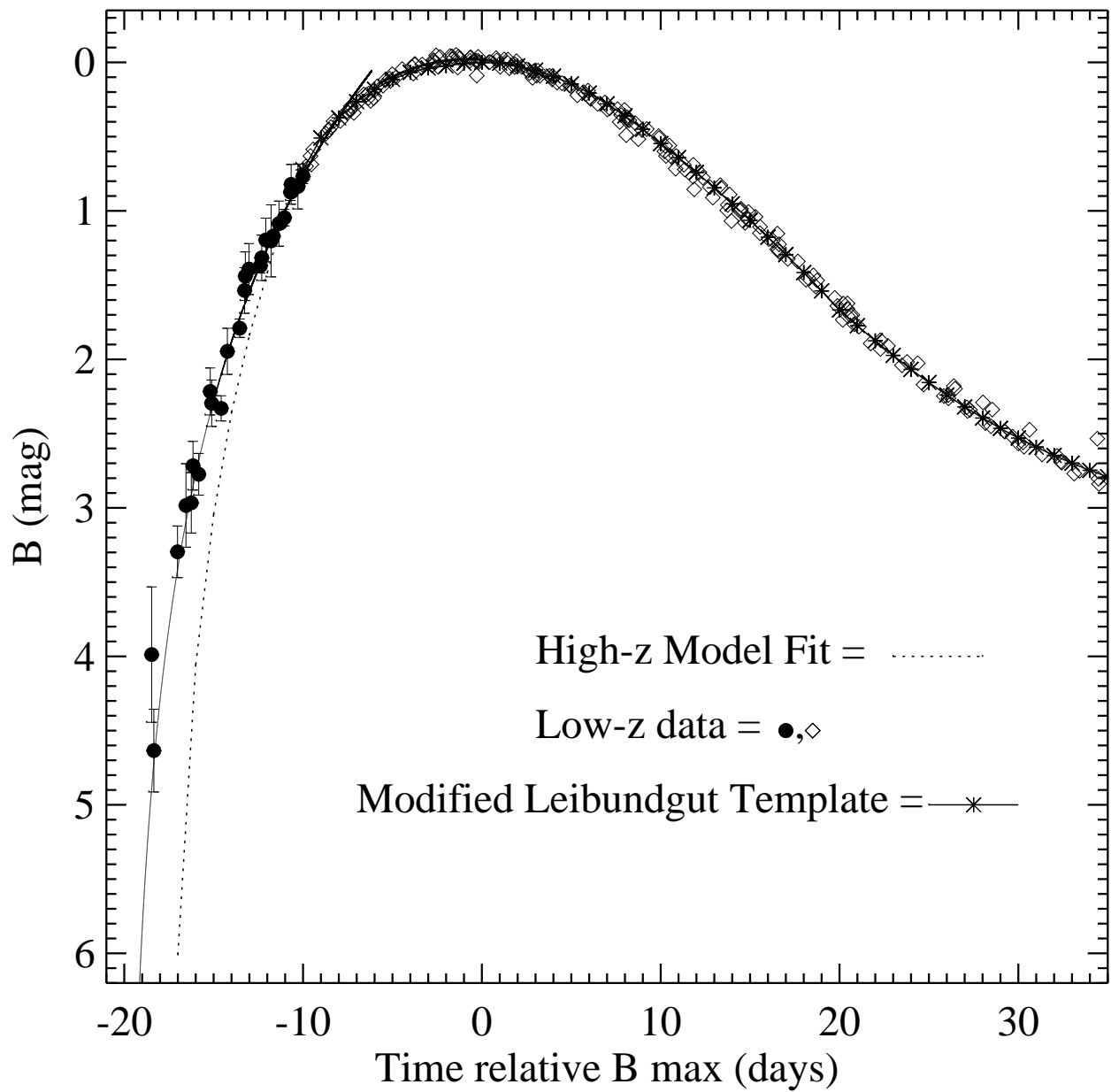
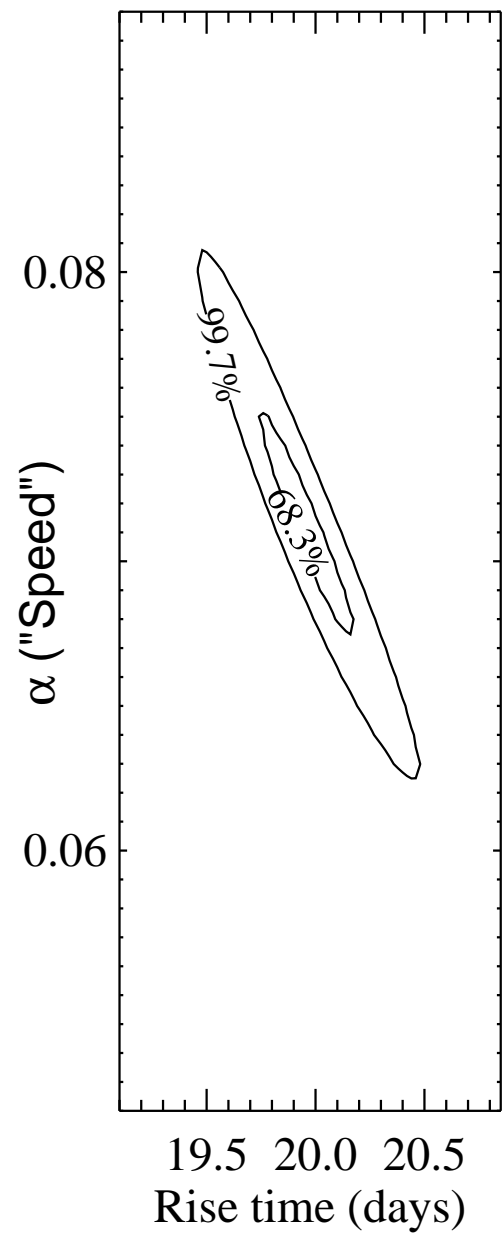
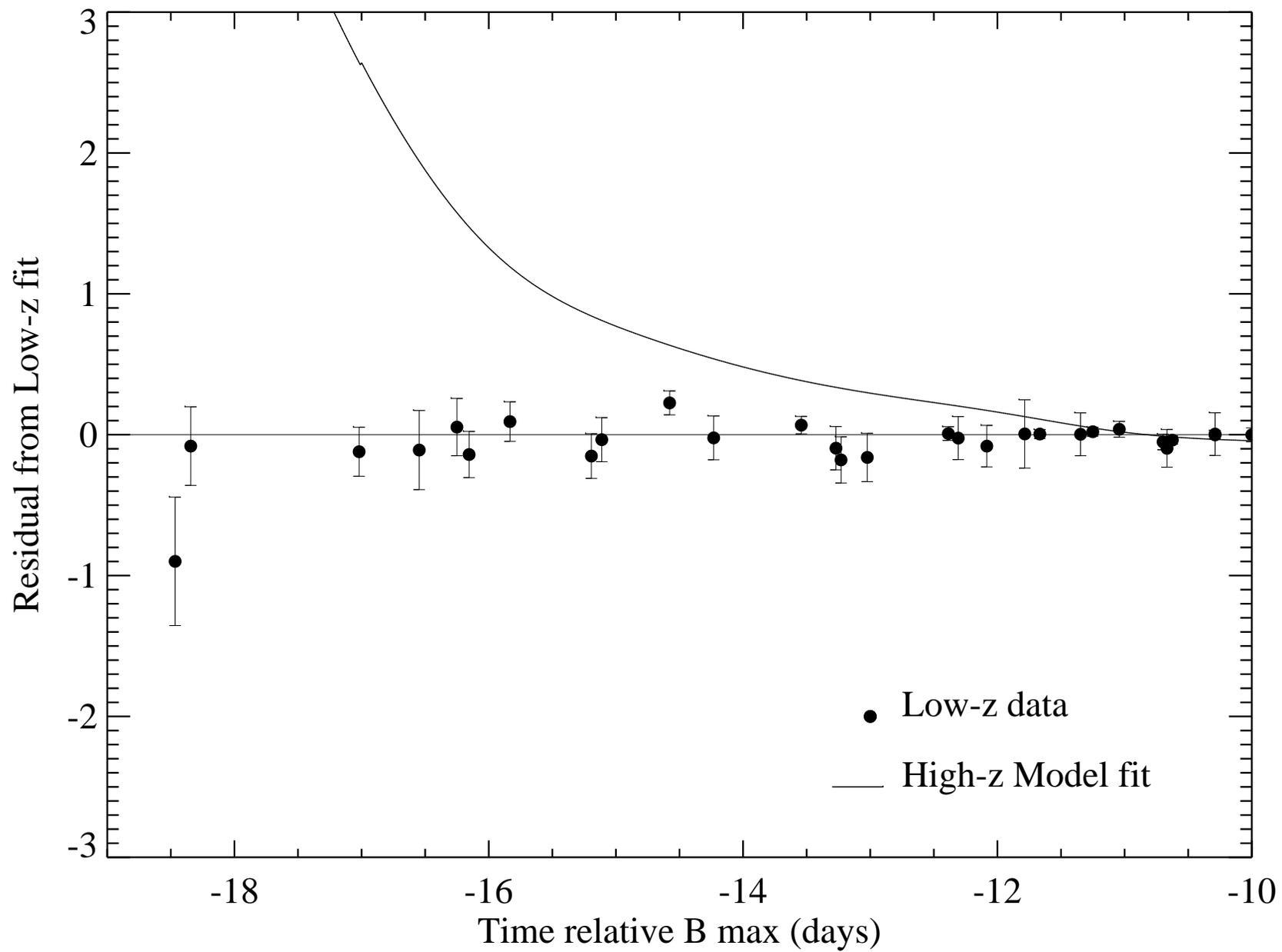
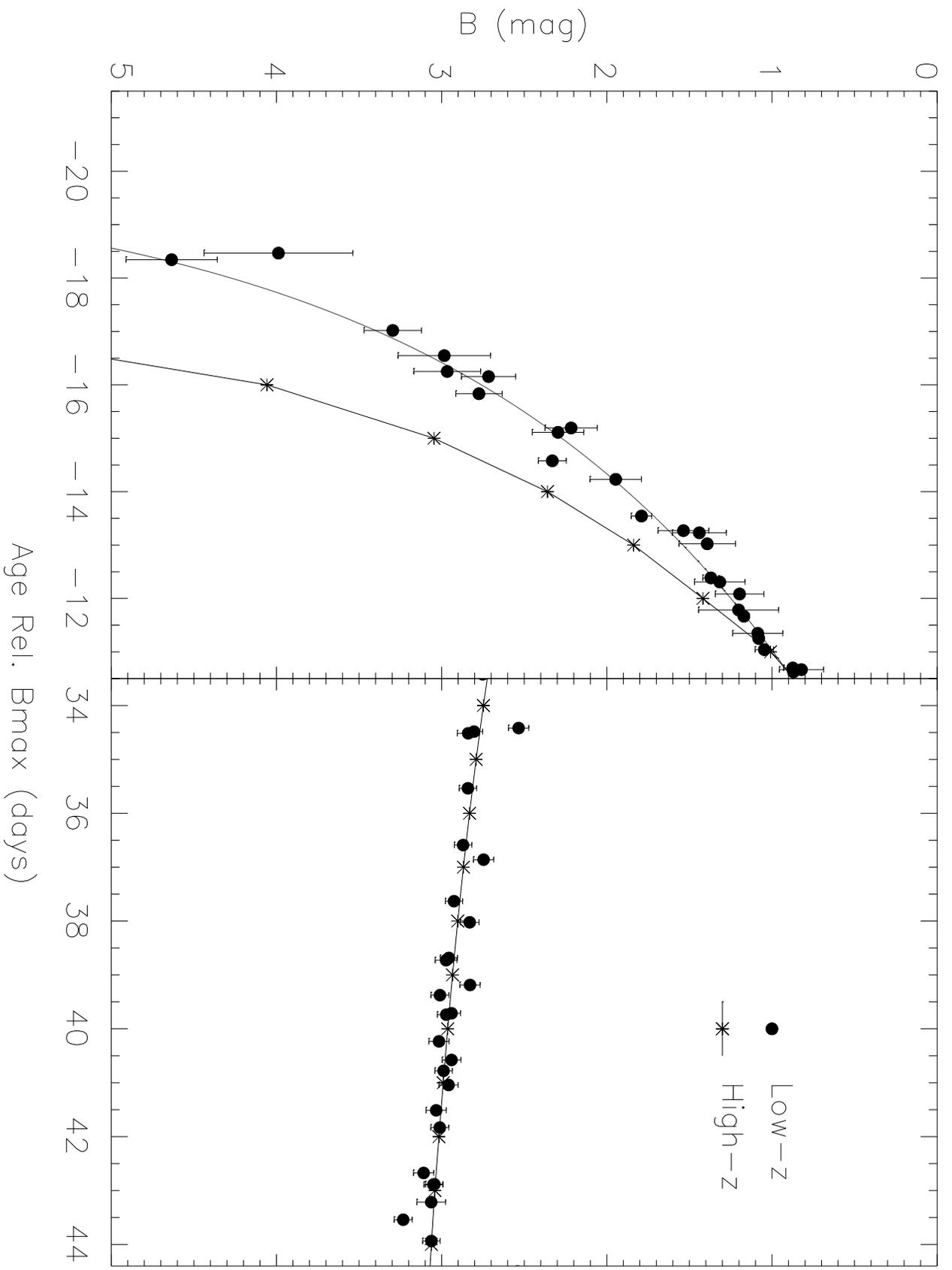
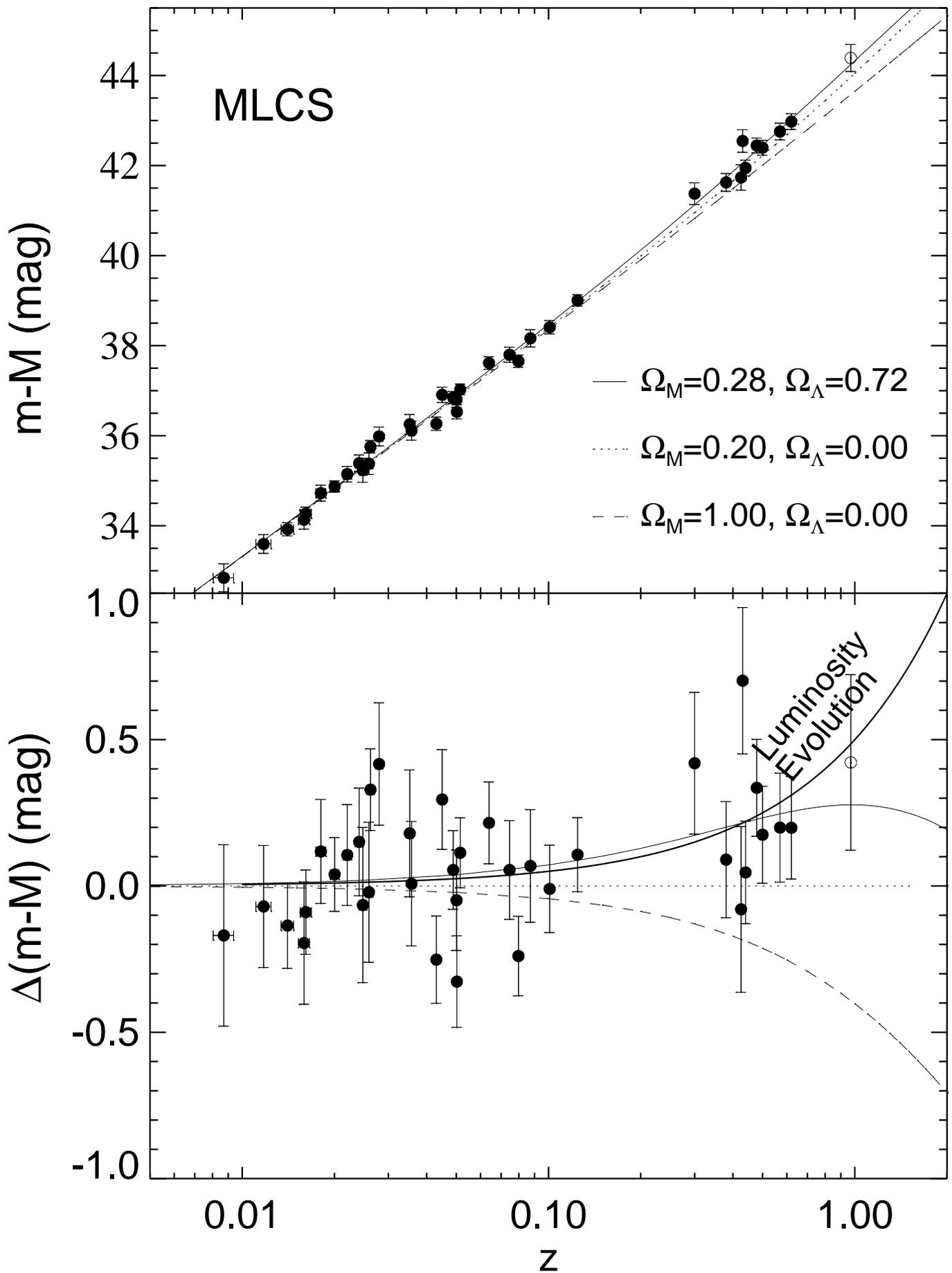
\end{document}